\documentstyle[12pt]{article}
\newcommand{\slp}{\raise.15ex\hbox{$/$}\kern-.57em\hbox{$\partial$}}
\newcommand{\sla}{\raise.15ex\hbox{$/$}\kern-.57em\hbox{$a$}}
\newcommand{\slA}{\raise.15ex\hbox{$/$}\kern-.57em\hbox{$A$}}
\newcommand{\slB}{\raise.15ex\hbox{$/$}\kern-.57em\hbox{$B$}}
\newcommand{\slb}{\raise.15ex\hbox{$/$}\kern-.57em\hbox{$b$}}
\newcommand{\slW}{\raise.15ex\hbox{$/$}\kern-.57em\hbox{$W$}}
\newcommand{\dA}{\sqcup\!\!\!\!\!\sqcap}
\newcommand{\be}{\begin{equation}}
\newcommand{\ee}{\end{equation}}
\newcommand{\bear}{\begin{eqnarray}}
\newcommand{\ear}{\end{eqnarray}}

\renewcommand{\theequation}{\arabic{section}.\arabic{equation}}

\begin{document}
\begin{titlepage}
\begin{flushright}
HD--THEP--97--19\\
\end{flushright}
\bigskip
\begin{center}
{\bf\Large DECOUPLED PATH INTEGRAL FORMULATION}\\
\vspace{.3cm}
{\bf\Large OF CHIRAL QCD$_2$ WITH $a_{JR}=2$}\\
\vspace{1cm}
{\bf R. L. P. G. Amaral and L. V. Belvedere}\\
\medskip
{\it Instituto  de F\'{\i}sica}\\
{\it Universidade Federal Fluminense}\\
{\it Av. Litor\^anea ,S/N, Boa Viagem, Niter\'oi, CEP. 24210-340}\\
{\it Rio de Janeiro - Brasil}\\
\bigskip
{\bf K. D. Rothe}\\
\medskip
{\it Institut  f\"ur Theoretische Physik}\\
{\it Universit\"at Heidelberg}\\
{\it Philosophenweg 16, D-69120 Heidelberg}\\
\bigskip
{\bf F.G. Scholtz}\\
\medskip
{\it Institute of Theoretical Physics}\\
{\it University of Stellenbosch}\\
{\it Stellenbosch 7600, South Africa}
\end{center}
\vspace{2.0cm}
\begin{abstract}
We analyse the BRST constraints and corresponding Hilbert-space
structure of chiral QCD$_2$ in the decoupled formulation
for the case of the Jackiw-Rajaraman parameter $a=2$.
We show that despite formal similarities this theory is not equivalent
to QCD$_2$, and that its extension to $U(N)$ does
not lead to an infinite vacuum degeneracy.
\end{abstract}
\end{titlepage}
\newpage
\section{Introduction}
The recent formulation of QCD$_2$ as a perturbed Wess-Zumino-Witten
(WZW) theory \cite{RS},\cite{AA1}, \cite{AA2} has provided some interesting
insight into structural aspects of the theory \cite{AA1}, 
\cite{CRS}, \cite{CR},
\cite{AR}. In the so-called non-local decoupled formulation
\cite{AA1} the corresponding (``enlarged'') Hilbert space is generated
from an effective partition function given
in terms of the direct product of non-interacting fermion
and ghost sectors, as well as a ``massive'' interacting
sector. The physical Hilbert subspace is obtained by imposing BRST conditions
on the states. As was shown in ref. \cite{AR}, these conditions
correspond in the abelian $U(1)$ case (vector Schwinger model) to the familiar
Lowenstein-Swieca \cite{LS} conditions requiring that the
longitudinal part of the current annihilate the
physical states. Analogous conditions have been obtained
by Bojanovsky et al. \cite{B} for the
case of the chiral Schwinger model \cite{JR,GRR}.
A corresponding analysis of this anomalous model (chiral QCD$_2$) is, however,
lacking. The purpose of this paper is to fill this gap. We shall in
particular concentrate on the case
of the Jakiw-Rajaraman parameter \cite{JR} $a$ taking the
value $a=2$, for which the chiral Schwinger model has
been claimed \cite{CW}
to be equivalent to the Schwinger model. We shall reexamine
this question in the context of chiral QCD$_2$ and show that,
like in the abelian case \cite{Bel}, this equivalence does
not exist.

In section 2 we show for $a=2$ that the effective Lagrangian
obtained in \cite{B} just corresponds to the abelian counterpart of
the non-local
decoupled formulation of chiral QCD$_2$, and that the conditions
imposed in \cite{B} on the physical states correspond to the BRST
conditions one systematically derives in the corresponding
non-abelian case. This will also serve to streamline
the presentation of ref. \cite{B}.

In section 3 we then discuss why chiral QCD$_2$ for $a=2$ is not
equivalent to QCD$_2$ by examining
in more detail its physical Hilbert space. Section 4 contains
our conclusions and some general remarks on the structure of the
physical Hilbert space. Considerations showing that the BRST
symmetry operating in the ``massive'' sector does not imply restrictions
on the physical states are relegated to the appendix.

\section{BRST constraints of chiral QCD$_2$}
\setcounter{equation}{0}
Our starting point is the (Minkowski) partition function
of chiral QCD$_2$, with left-handed fermions coupled to
a $SU(N)$ gauge field:
\be\label{2.1}
Z=\int{\cal D} A_\mu\int{\cal D}\psi^{(0)}_1{\cal D}\psi
_1^{(0)\dagger}\int {\cal D}\psi_2{\cal D}\psi_2^\dagger\, e^{\,i\,S[A,\psi,
\bar\psi]}\ee
with$^1$
\be\label{2.2}
S[A,\psi,\bar\psi]=\int d^2x\left\{
-\frac{1}{4}tr F^{\mu\nu}F_{\mu\nu}+
\psi_1^{\dagger(0)}i\partial_+\psi_1^{(0)}+\psi^\dagger_2
(i\partial_-+eA_-)\psi_2\right\}\ee
Parametrizing $A_\pm$ as follows
\be\label{2.3}
eA_+ =U^{-1}i\partial_+U,\ eA_-=Vi\partial_-V^{-1},\ee
the Yang-Mills action (\ref{2.2}) then can be written in the
two alternative forms \cite{RST}
\be\label{2.4}
S_{YM}[\Sigma]=\frac{1}{4e^2}\int tr\frac{1}{2}
[\partial_+(\Sigma i\partial_-\Sigma^{-1})]^2\ee
\be\label{2.5}
\qquad =\frac{1}{4e^2}\int tr\frac{1}{2}
[\partial_-(\Sigma^{-1}i\partial_+\Sigma)]^2\ee
with $\Sigma$ the gauge-invariant variable
\be\label{2.6}
\Sigma=UV.\ee
Making the change of variables
$A_+\to U,\quad A_-\to V$
as well as the chiral rotation,
\be\label{2.7}
\psi_2\to\psi_2^{(0)}:\psi_2^{(0)}=V^{-1}\psi_2,\ee
and taking due account of the Jacobians in the integration measure \cite{RS}
\cite{AA1}, we then arrive, following the procedure of
references \cite{AA1}, \cite{CRS}, at the partition function
\be\label{2.8}
Z=Z_F^{(0)}Z_{gh}^{(0)}\hat Z,\ee
where $Z^{(0)}_F$ is the partition function of free fermions,
\be\label{2.9}
Z_F^{(0)}=\int{\cal D}\psi^{(0)}{\cal D}\bar\psi^{(0)}e^{i
\int\bar\psi^{(0)}i\slp\psi^{(0)}},\ee
$Z_{gh}^{(0)}$ is the partition function of free ghosts
associated with the change of variables (\ref{2.3})
\be\label{2.10}
Z_{gh}^{(0)}=\int{\cal D}b^{(0)}_+{\cal D}c_+
^{(0)}e^{i\int tr b_+^{(0)}i\partial_-c_+^{(0)}}
\int{\cal D}b_-^{(0)}{\cal D}c_-^{(0)}e^{i\int tr b_-^{(0)}i
\partial_+c_-^{(0)}},\ee
and where
\be\label{2.11}
\hat Z=\int{\cal D}U{\cal D}Ve^{iS_{YM}[UV]}
e^{-iC_V\Gamma[UV]-i\Gamma[V]}e^{iS_{JR}}.\ee
Here $\Gamma[g]$ is the WZW functional \cite{WZW},
\be\label{2.12}
\Gamma[g]=\frac{1}{8\pi} \int d^2x tr{\partial_\mu g \partial^\mu g^{-1}}
+\frac{1}{12\pi} \int_{S_B} d^3x \epsilon^{ijk}
tr[\hat g^{-1} \partial_i \hat g \hat g^{-1}\partial_j \hat g
\hat g^{-1} \partial_k \hat g]\ee
and
\be\label{2.13}
S_{JR}= \frac{a}{2}\frac{e^2}{4\pi}\int d^2x\ tr [A_+A_-] =
-\frac{a}{2}\frac{1}{4\pi}\int d^2x tr
[(U^{-1}\partial_+U)(V\partial_-V^{-1})] .
\ee
The presence of the last factor in (\ref{2.11}) reflects
the usual regularization ambiguity, with $a$ the Jackiw-Rajaraman parameter
\cite{JR}.

Note that because of the presence of the factor $exp(iS_{JR})$, our change of
variables did not result in a decoupling of the fields,
unlike in the case of QCD$_2$. Nevertheless, for $a=2$ a decoupling of these
fields is easily achieved.
Indeed, making use of the Polyakov-Wiegmann identity
\cite{PW},
\be\label{PW}
\Gamma[gh]=\Gamma[g]+\Gamma[h]+\frac{1}{4\pi}\int tr
[(g^{-1}\partial_+g)(h\partial_-h^{-1})],\ee
we obtain from (\ref{2.8})
\be\label{Za}
Z\,=\,Z_{_{_F}}^{^{(0)}}\,Z_{_{_{gh}}}^{^{(0)}}\,\int {\cal D}U\,{\cal D}V\,
e^{\,i\,S_{_{_{YM}}}[UV]}\,e^{\,-\,i\,(C_{_V}\,+\,\frac{a}{2})\,\Gamma[UV]}\,
e^{\,i\,\frac{a}{2}\,\Gamma[U]}\,e^{\,i\,(\frac{a}{2}\,-\,1)\,\Gamma [V]}\,.
\ee
Unlike the case of of $QCD_2$, the transformations (\ref{PW})
and (\ref{2.7}) have not led to a decoupling of the fields. However,
for $ a = 2 $, (\ref{Za}) reduces to the decoupled partition function
\be\label{2.17}
Z=Z_{_{_F}}^{^{(0)}}\,Z_{_{_{gh}}}^{^{(0)}}\,\Bigl (\,\int\,{\cal D}U\,
e^{\,i\,\Gamma[U]}\,\Bigr )\,\Bigl (\,\int {\cal D}\Sigma\,
e^{\,i\,S_{_{_{YM}}}[\Sigma]}\,e^{\,-\,i\,(1\,+\,C_{_V})\,\Gamma[\Sigma]}\,
\Bigr )\,.
\ee
Except for the factor $ \exp \{\,i\,\Gamma[U]\,\} $, which appears
to play merely a spectator role, this is the partition function
of QCD$_2$ in the decoupled formulation \cite{AA1,AA2,CRS,CR,RST}.
 As we shall see, however, the
apparently decoupled field $U$ plays an
important role in the analysis of the physical Hilbert space$^2$.

Let us check where we stand by specializing to the abelian
case. Introducing in that case the parametrization
\be\label{2.18}
U=e^{-ie(\chi+\lambda)},\ V=e^{-ie(\chi-\lambda)}\ee
and noting that $C_V=0$ for $U(1)$, we obtain from (\ref{2.17})
($Z_{gh}^{(0)}=1$),
\[Z=\int{\cal D}\phi{\cal D}\chi{\cal D}\lambda e^{i\int{\cal
L}^{(Bos)}_{U(1)}}\]
with
\be\label{2.19}
{\cal L}^{(Bos)}_{U(1)}=\frac{1}{2}(\partial_\mu\phi)^2+\frac{1}{2}
(\dA\chi)^2-\frac{1}{2}m^2(\partial_\mu\chi)^2+\frac{1}
{2}(\partial_\mu\varphi)^2,\ee
where we have set $\chi+\lambda=\frac{2\sqrt{\pi}}{e}\varphi$, $ m^2
= 4 e^2 / \pi $ and have
made use of bosonization formula
\be\label{2.20}
\bar\psi^{(0)}i\slp\psi^{(0)}\to\frac{1}{2}
(\partial_\mu\phi)^2.\ee
The Lagrangian (\ref{2.19}) is just the one obtained in ref.
\cite{B} for $a=2$ after a suitable redefinition of the fields.
The appearance of fourth-order derivative
terms in the Lagrangian (\ref{2.19})
is already evident from (\ref{2.4}) and (\ref{2.5}).

In order to obtain a reduction to second-order derivatives
we introduce an auxiliary Lie-algebra-valued field $E$ or $E'$,
depending on which form of the Yang-Mills action, eq. (\ref{2.4})
or (\ref{2.5}), we choose to work with:
\bear
e^{iS_{YM}[\Sigma]}&=&\int{\cal D} Ee^{i\int tr[\frac{1}{2} E^2
+\frac{E}{2e}\partial_+(\Sigma i\partial_-\Sigma^{-1})]}\label{2.21}\\
&=&\int{\cal D} E'e^{i\int tr[\frac{1}{2} {E'}^2
+\frac{E'}{2e}\partial_-(\Sigma^{-1} i\partial_+\Sigma)]}.\label{2.22}
\ear
Making the change of variable \cite{CR}
\be\label{2.23}
\partial_+E=\left(\frac{1+C_V}{2\pi}\right)
e\beta^{-1}i\partial_+\beta\ee
and
\be\label{2.24}
\partial_-E'=\left(\frac{1+C_V}{2\pi}\right)
e\beta'i\partial_-{\beta'}^{-1}\ee
one arrives at two alternative  ``non-local''forms of the
decoupled partition function:
\bear\label{2.25}
Z&=&Z_F^{(0)}Z^{(0)}_{gh}\int{\cal D}
\tilde\Sigma{\cal D}\beta{\cal D}U e^{-i\left(\frac{1+C_V}
{2\pi}\right)^2e^2\int\frac{1}{2}tr\left[\partial_+^{-1}
(\beta^{-1}i\partial_+\beta)\right]^2}\nonumber\\
&&\times e^{-i(1+c_V)\Gamma[\tilde\Sigma]}
e^{i\Gamma[\beta]}e^{i\Gamma[U]}
 \int{\cal D}\hat b_-^{(0)}{\cal D}\hat c^{(0)}_-
e^{i\int tr\hat b_-^{(0)}i\partial_+\hat c^{(0)}_-}
\ear
\bear\label{2.26}
&=&Z_F^{(0)}Z^{(0)}_{gh}\int{\cal D}
\tilde\Sigma{\cal D}U{\cal D}\beta' e^{-i\left(\frac{1+C_V}
{2\pi}\right)^2e^2\int\frac{1}{2}tr\left[\partial_-^{-1}
(\beta'i\partial_-{\beta'}^{-1})\right]^2}\nonumber\\
&&\times e^{-i(1+C_V)\Gamma[\tilde\Sigma']}
e^{i\Gamma[\beta']}e^{i\Gamma[U]}
\int{\cal D}\hat b_+^{(0)}{\cal D}\hat c^{(0)}_+
e^{i\int tr\hat b^{(0)}_+i\partial_-\hat c^{(0)}_+}
\ear
where the changes of variable
\be\label{2.27}
\tilde\Sigma=\beta\Sigma,\quad\tilde\Sigma'=\Sigma\beta'\ee
have been made and use has been made of
the Polyakov-Wiegmann identity (\ref{PW}). Again, aside from (the all important factor)
exp$ i\Gamma[U]$, this is just the QCD$_2$ partition
function before gauge fixing.

As before let us check where we stand by considering the special case of
an abelian $U(1)$ group. As one easily verifies, one has in this case
$\beta=\beta'$ (and hence $\tilde \Sigma ' = \tilde \Sigma)$.
 Parametrizing $\beta$, $\tilde\Sigma$ and $U$ as follows
\be\label{2.28}
\beta=e^{-2i\sqrt\pi\sigma},\quad \tilde\Sigma=e^{-2i\sqrt\pi\eta},
\quad U=e^{-2i\sqrt\pi\varphi}\ee
expressions (\ref{2.25})
and (\ref{2.26}) reduce to
\be\label{2.31}
Z=\int {\cal D}\phi{\cal D}\varphi{\cal D}\sigma{\cal D}\eta
e^{i\int\tilde{\cal L}_{U(1)}^{(Bos)}}\ee
with
\be\label{2.32}
\tilde{\cal L}^{(Bos)}_{U(1)}=\frac{1}{2}(\partial_\mu \phi)^2+\frac
{1}{2}(\partial_\mu \sigma)^2-\frac{1}{2}m^2\sigma^2-
\frac{1}{2}(\partial_\mu \eta)^2 + \frac{1}{2}(\partial_\mu \varphi)^2,\ee
where use has again been made of the bosonization formula
(\ref{2.20}). Expression (\ref{2.32}) is identical, after suitable
relabelling, with the one
obtained in ref. \cite{B} after a series of manipulations. Note
that $\eta$ is a negative metric (unphysical) field,
corresponding to the fact that $\tilde\Sigma (\tilde\Sigma^\prime)$
in eq. (\ref{2.25}) (eq. (\ref{2.26})) is a WZW field of negative level
$-(1+C_V)$. Except for the last term, the Lagrangian (\ref{2.32}) coincides
with that of the vector Schwinger model (VSM). In the VSM the gauge
invariance ensures that the field $ \varphi $ is a pure gauge
excitation and does not appear in the bosonized Lagrangian density
\cite{B,Bel}. However, in the anomalous chiral model the additional degree
of freedom $ \varphi $ is a dynamical field and its
presence ensures the existence of fermions in the asymptotic states \cite{Bel}.

Despite the factorized form (\ref{2.25}) (or (\ref{2.26}))
of the partitition function, the physical Hilbert space is not just
the direct product of decoupled sectors, but is restricted
by a set of BRST condition, which can be derived from first
principles, following the procedure of ref. \cite{Ba}.
These conditions ensure in particular that negative
norm states associated with $\tilde\Sigma(\tilde{\Sigma'})$
are absent in $H_{phys}$. Following
the steps outlined in refs. \cite{CR} and \cite{RST},
one thus systematically discovers the existence of three nilpotent
transformations. They are associated with the changes of variable
1) $A_-\to V$, 2) $A_+\to U$ and 3a) $E\to\beta$, or
3b) $E'\to\beta'$, respectively. According to the discussion in
the appendix only the first two imply non-trivial restrictions
on the physical states. They
are derived from (\ref{2.1})\cite{RST} and represent nilpotent
symmetry transformations of
the local partition function (\ref{2.8}):
\bear\label{2.34}
1)\quad &&\delta U=0,\quad \delta V^{-1}V=c_+^{(0)}\nonumber\\
&&\delta\psi_1^{(0)}=0,\quad \delta\psi^{(0)}_2=
c_+^{(0)}\psi_2^{(0)}\nonumber\\
&&\delta c_-^{(0)}=0,\quad \delta c_+^{(0)}=\frac{1}{2}
\{c_+^{(0)},c_+^{(0)}\}\nonumber\\
&&\delta b_-^{(0)}=0\nonumber\\
\delta b_+^{(0)}&=&\frac{1}{4e^2}\Sigma^{-1}[\partial_+^2
(\Sigma i\partial_-\Sigma^{-1})]\Sigma\nonumber\\
&-&\frac{1+C_V}{4\pi}
\Sigma^{-1}i\partial_+\Sigma
+\psi_2^{(0)}\psi_2^{(0)\dagger}+\{b^{(0)}_+,c_+^{(0)}\}
\ear
and
\bear\label{2.35}
2)\quad&&\delta V=0,\quad\delta U U^{-1}=c_-^{(0)}\nonumber\\
&&\delta\psi_1^{(0)}=c^{(0)}_-\psi^{(0)}_1,\quad\delta\psi_2^{(0)}=0\nonumber\\
&&\delta c_+^{(0)}=0,\quad \delta c_-^{(0)}=\frac{1}{2}
\{c_-^{(0)},c_-^{(0)}\}\nonumber\\
&&\delta b^{(0)}_+=0\nonumber\\
\delta b_-^{(0)}&=&\frac{1}{4e^2}\Sigma[\partial_-^2(\Sigma^{-1}
i\partial_+\Sigma)]\Sigma^{-1}\nonumber\\
&-&\frac{1+C_V}{4\pi}\Sigma i\partial_-
\Sigma^{-1}
+\frac{1}{4\pi}Ui\partial_-U^{-1}+\{b_-^{(0)},c_-^{(0)}\}.\ear
The BRST transformations associated with the changes of variable 3a) and 3b)
are of exactly the same form as in the case of $QCD_2$ \cite{CR}.
As we show in the appendix, they however imply no restriction on the
physical states.

Going through the usual Noether construction, it is straightforward
to show that these symmetry transformations imply the existence
of the following conserved charges:
\be\label{2.39}
Q_\pm=\int dx^1tr c_\pm^{(0)}(\Omega_\pm
-\frac{1}{2}\{b_\pm^{(0)},c_\pm^{(0)}\})\ee
where
\be\label{2.41}
\Omega_\pm=\delta b_\pm^{(0)}.\ee
Canonical quantization (see later) shows that
$\Omega^a_\pm = trt^a\Omega_\pm$
are weakly first-class operators. As a result, $Q_\pm$ are nilpotent
 \cite{He}.

The nilpotency of the  charges, together
with the condition that they annihilate the physical
states, implies that $\Omega_\pm$ must vanish on such states.
Let us express $\Omega_\pm$ in terms of the variables
of the non-local formulation. Following ref. \cite{CR}
we make use
of the equation of motion
\bear\label{2.42}
E&=&-\frac{1}{2e}\partial_+(\Sigma i\partial_-\Sigma^{-1})
\nonumber\\
E'&=&-\frac{1}{2e}\partial_-(\Sigma^{-1} i\partial_+\Sigma)
\ear
as well as of (\ref{2.23}) and (\ref{2.24}) in the expressions
(\ref{2.34}) and (\ref{2.35}) for
$\delta b_\pm^{(0)}$, which allows us to reduce $\Omega_\pm$
to the simple form
\bear\label{2.43}
&&\Omega_+=-\frac{1+C_V}{4\pi}\tilde\Sigma^{-1}i\partial_+\tilde\Sigma
+\psi_2^{(0)}\psi_2^{(0)\dagger}+\{
b_+^{(0)},c_+^{(0)}\}\nonumber\\
&&\Omega_-=-\frac{1+C_V}{4\pi}\tilde\Sigma'i\partial_-\tilde{\Sigma'}
^{-1}+\frac{1}{4\pi}Ui\partial_-U^{-1}
+\{b_-^{(0)},c_-^{(0)}\}.\ear
Let us check once more where we stand by
considering the abelian case. As before we have in this case
$\beta=\beta'$, $\tilde\Sigma=\tilde\Sigma'$. With the
parametrizations (\ref{2.20}) and (\ref{2.28}),
the conditions $\Omega_\pm\approx 0$ reduce to
\bear\label{2.44}
&&\Omega_+=-\frac{1}{2\sqrt\pi}\partial_+\phi-\frac{1}{2\sqrt\pi}
\partial_+\eta\approx 0,\nonumber\\
&&\Omega_-=+\frac{1}{2\sqrt\pi}\partial_-\eta-\frac{1}{2\sqrt\pi}
\partial_-\varphi\approx0,\ear
which may be summarized by
\be\label{2.45}
\Omega_\mu=-\frac{1}{4\sqrt\pi}(\partial_\mu+\epsilon_{\mu
\nu}\partial^\nu)\phi-\frac{1}{4\sqrt\pi}(\partial_\mu-\epsilon_{\mu\nu}
\partial^\nu)\varphi-\frac{1}{4\sqrt\pi}\epsilon_{\mu\nu}\partial^\nu
\eta\approx0.\ee
Except for a trivial relabelling,
these are precisely the conditions obtained in ref. \cite{B}
for the case of the chiral Schwinger model, from another point
of view.

\section{The physical Hilbert space}
\setcounter{equation}{0}

In order to get a more detailed understanding of the implications of the
constraints $\Omega_\pm \approx 0$ on the physical states, we must go over to
phase space.
The canonical quantization proceeds as in \cite{CR}.
In terms of phase-space variables, $\Omega_\pm$ and
  then takes the form
\be\label{3.1}
\Omega_+=J_+(\tilde\Sigma)+\psi_2^{(0)}\psi_2^{(0)\dagger}
+\{b_+^{(0)},c_+^{(0)}\}\ee
\be\label{3.2}
\Omega_-=J_-(\tilde\Sigma')+J_-(U)+\{b_-^{(0)},c_-^{(0)}\}\ee
where (see ref.\cite{CR} for notation and more details;
the superscript ``T'' denotes ``transpose'')
\be\label{3.5}
J_+(\tilde\Sigma)=-i\hat\Pi^T_{\tilde\Sigma}\tilde\Sigma-
\frac{1+C_V}{4\pi}\tilde\Sigma^{-1}i\partial_1\tilde\Sigma\ee
\be\label{3.6}
J_-(\tilde\Sigma')=i{\tilde\Sigma}'\hat\Pi^T_{\tilde\Sigma'}
+\frac{1+C_V}{4\pi}\tilde\Sigma'i\partial_1\tilde{\Sigma'}
^{-1}\ee
\be
J_-(U) = iU\hat\Pi^T_U - \frac{1}{4\pi} Ui\partial_1 U^{-1}\ee
The WZW-currents $J_+(\tilde\Sigma),J_-(\tilde\Sigma')$ and $J_-(U)$
satisfy a Kac-Moody algebra$^3$
$(J_\pm^a=tr t^aJ_\pm)$
\bear\label{3.9}
&&\{J_\pm^a (h(x)),J_\pm^b(h(y))\}_P=-f^{abc }J_\pm(h(x))
\delta(x^1-y^1)+\frac{\kappa}{2\pi}\partial_1\delta^{ab}\delta
(x^1-y^1)\nonumber\\
&&\{J_+^a(h(x)),J^b_-(h(y))\}_P=0\ear
of level $\kappa=-(1+C_V),-(1+C_V)$ and 1, respectively.

We next show that the fields $ \psi^{(0)}_1,\psi_2$ and $A_\mu$
commute with the operators (\ref{3.1})-(\ref{3.2}),
and hence represent (physical) observables of the theory.
To this end we first rewrite these fields in terms of the fields
of the decoupled
non-local formulation $U,
\tilde\Sigma,\beta$ and their canonical conjugates :
\be\label{3.10}
\psi_2=V\psi_2^{(0)}=U^{-1}\Sigma\psi_2^{(0)}=
U^{-1}\beta^{-1}\tilde\Sigma\psi_2^{(0)}\ee
\be\label{3.11}
eA_+=U^{-1}i\partial_+U=4\pi J_+(U)\ee
\bear\label{3.12}
eA_-&=&Vi\partial_-V^{-1} =
(U^{-1}\beta^{-1}\tilde\Sigma)i\partial_-(\tilde\Sigma^{-1}\beta U)\\
&=&-\frac{4\pi}{1+C_V} (\beta U)^{-1}J_-(\tilde\Sigma)(\beta U)
-U^{-1}J_-(U)U + U^{-1}(\beta^{-1} i\partial_- \beta )U\nonumber\ear
or in terms of $\tilde\Sigma'$,
\bear
eA_- &=& U^{-1}\tilde\Sigma'(\beta'^{-1}i\partial_-\beta')\tilde\Sigma'^{-1}U
\nonumber\\
&-& \frac{4\pi}{1+c_V} U^{-1}J_-(\tilde\Sigma')U - U^{-1}J_-(U)U.
\ear

i) First of all, $\psi_1^{(0)}$ commutes with all
constraints, since they do not involve $\psi_1^{(0)}$.
Hence, {\it $\psi_1^{(0)}$ is physical}.

ii) We have (different sectors commute with each other)
\be\label{3.15}
\{J_+^a(\tilde\Sigma(x)),U^{-1}\beta^{-1}\tilde\Sigma
\psi_2^{(0)}(y)\}_P=iU^{-1}{\beta^{-1}}\tilde\Sigma t^a\psi^{(0)}_2(x)
\delta(x^1-y^1)\ee
and
\be\label{3.16}\{tr(t^a\psi_2^{(0)}\psi_2^{(0)\dagger}
(x)),U^{-1}\beta^{-1}\tilde\Sigma\psi_2^{(0)}(y)\}_P
=-iU^{-1}\beta^{-1}\tilde\Sigma t^a\psi_2^{(0)}(x)\delta(x^1-y^1)\ee
Hence $\{\Omega_+(x),\psi_2(y)\}_P=0$.
In a similar way one verifies that
\newline
$\{\Omega_-(x),\psi_2(y)\}_P=0$.
Hence {\it $\psi_2$ is physical}.

iii) $eA_+=U^{-1}i\partial_+U$ evidently commutes with
$\Omega_+^a$. As for $\Omega_-^a$ we have
\be\label{3.17}
\{\Omega_-^a(x),A_+^b(y)\}_P=4\pi\{J_-^a(U(x)),J^b_+(U(y))\}_P=0\ee
Hence {\it $A_+$ is physical}.

iv) Similar considerations show that $A_-$ also commutes with
$\Omega_\pm$.The vanishing of the Poisson bracket
$\{\Omega_+^a(x),A_-(y)\}_P$ follows from the commutativity of $J_+$ with
$J_-$.
Furthermore, making use of the Poisson brackets
\bear
\{J_-^a(h(x)),h(x)\}_P= -i(t^ah(x))\delta(x^1-y^1)\nonumber\\
\{J_-^a(h(x)),h^{-1}(y)\}_P= i(h^{-1}(x)t^a)\delta(x^1-y^1)
\ear
and
\bear
\{J_-^a(h(x)),h^{-1}Qh(y)\}_P
&=& i(h^{-1}\left[ t^a,Q\right]h)\delta(x^1-y^1)\nonumber\\
&+& h^{-1}t^bh(x)\{J_-^a(h(x)),Q^b(y)\}_P
\ear
one finds that all contributions (including the central terms) to
$\{\Omega_-^a(x),A_-(y)\}_P$ cancel, so that {\it $A_-$ is physical}, as
well.

Summarizing, the BRST conditions just tell us, that the physical Hilbert
space is constructed from $\{U,V,\psi^{(0)}_1, \psi^{(0)}_2\}$
as combinations of the basic fields of the original Lagrangian,
as expected.

\section{Conclusions and Final Remarks}

We have presented a decoupled path integral formulation of the
anomalous chiral QCD$_2$ for the case $a = 2$. In this section we
summarize our conclusions and also make
some general remarks on the structure of
the physical Hilbert space of the anomalous chiral model, which are
crucial to distinguish it from the gauge invariant case.

In order to
construct the Hilbert space associated with the
Wightmam functions that define the theory we only use the field algebra
generated from the intrinsic irreducible set of field operators of
the theory. The local field algebra $ \Im $ intrinsic to the chiral model is
generated from the {\it irreducible set of field
operators} $ \{\psi^{^{(0)}}_{_1}, \psi_{_2}, A_\mu\} $ \cite{Bel,MPS}. These
field operators constitute the intrinsic mathematical description of
the model and serve as a kind of building material in terms of which the theory
is formulated and whose Wightman functions define the model. The field
algebra $ \Im $ is represented on the Hilbert
space $ {\cal H} $.

The present techniques used to treat QFT models, such as bosonization
and Faddeev-Popov prescription, require the use of a larger field algebra,
which includes non-observable bosonic fields as well as ghosts. The
partition function of the resulting effective theory, for example in
the local decoupled formulation, is given in terms
of the set of fields  $ \{\psi^{^{(0)}}_{_1}, \psi^{^{(0)}}_{_2}, U, V,
gh\} $, which defines an enlarged redundant
field algebra $ \Im^e $. The field algebra $\Im^e$ is represented on the
Hilbert space $ {\cal H}^e $, which is the direct
product of decoupled sectors appearing in the effective partition
function (\ref{2.17}). The
fields $ \{\psi^{^{(0)}}_{_1}, \psi^{^{(0)}}_{_2}, U, V, gh\} $ should not
be considered as
elements of the field algebra intrinsic to
the model.

The field algebra $ \Im $ is a subalgebra of $ \Im^e $. Not all
functional of the
fields $ \{\psi^{^{(0)}}_{_1}, \psi^{^{(0)}}_{_2}, U, V, gh\} $  belong to the
algebra $ \Im $ of the local
fields, nor all vectors of $ {\cal H}^e $ belong to the state
space $ {\cal H} $ of the model. The set of states
corresponding to the largely redundant field algebra $ \Im ^e $, contains
elements which
are not intrinsic to the model. Of course, the Hilbert space $ {\cal H} $ is a
subspace of $ {\cal H}^e $.

The physical states of ${\cal H}$ are required to satisfy the BRST
subsidiary conditions
$$ {\cal Q}_{\pm} \vert \Psi \rangle = 0\,\,\,,\,\,\,\forall\,
\vert \Psi \rangle\,\in\,{\cal H}_{_{_{phys}}}\,, $$
which ensure that unphysical fields violating norm-positivity
appear in the physical subspace only as zero-norm combinations. This
condition also ensures that the physical Hilbert space cannot be
decomposed as a direct product of decoupled sectors appearing in the
effective partition function (\ref{2.25}),(\ref{2.26}).
As a consequence, the dependence on
the ``{\it apparently decoupled field}'' $ U $ cannot be factorized from the
completion of states. In this way, the partition function of the
anomalous chiral model cannot be factorized as a direct product of a coset
model and a massive model \cite{AR}, and thus cannot be identified
with the partition function of QCD$_2$.

The algebra of the physical operators $ \Im_{_{phys}} $ must be
identified as the subalgebra $\Im^e$ of $ \Im $ which obeys the
constraint condition in a proper Hilbert space completion of
states. As we have shown in the previous section, it is a peculiarity of the
anomalous chiral model that the algebra $ \Im_{_{phys}} \equiv \Im $, since
all operators belonging to the intrinsic  field algebra $ \Im $ commute
with the BRST constraints and thus represent physical observables of the
theory. In analogy with the abelian case and in
agreement with the conclusions of refs.\cite{B,Bel,GR}, this is
expected in the anomalous case since the
quantum theory has lost the local gauge invariance. In contrast to
the QCD$_2$, the field algebra $ \Im $ is {\it physical} and is represented
in the state space $ {\cal H}_{_{phys}} \equiv {\cal H} $.

The fact that the irreducible set of field
operators $ \{\psi^{^{(0)}}_{_1}, \psi_{_2}, A_\mu\} $ represents
physical observables of the chiral QCD$_2$ is a peculiar structural aspect
of the anomalous theory which allows for two
isomorphic formulations, i.e., the gauge invariant and gauge noninvariant
formulations \cite{Bel,GR}. This property of the algebra of
observables enables a basic structural distinction between the
anomalous chiral theory and the gauge invariant one and also
corroborates to support the conclusion that {\it in
analogy with the abelian case (chiral QED$_2$) \cite{Bel}, chiral QCD$_2$
for the JR parameter $ a = 2 $ is not equivalent to QCD$_2$.}

The structural properties of the model must be analyzed taking a
careful control on the Hilbert space associated with the Wightman
functions that provide a representation of the field algebra
generated from the intrinsic irreducible set of field operators of
the model. This rigorous control on the construction of the Hilbert
space together with the BRST constraints, constitute
the necessary and sufficient requirements to identify correctely the
physical operators in the effective decoupled
formulation of the anomalous theory.

In order to show that by relaxing this rigorous control on the
construction of the physical Hilbert space of the anomalous theory
some misleading conclusions can arise, as for example the existence
of an infinite number of states which are degenerate in energy with
the vacuum, it is interesting to examine the effect of extending
the group $SU(N)$ to $U(N)$. In the case of a true gauge theory, we expect
in this case an infinite degeneracy of the ground state as it is
known to occur in the Schwinger model.

Let us first specialize to the $U(1)$ case,
that is, the chiral Schwinger model with $a = 2$. This particular case has
generated some confusion in the literature since
the operator algebra exhibits non-trivial and delicate features
\cite{Bel} which
might lead to misleading conclusions \cite{CW} about structural properties of
the model. As stressed in ref.\cite{Bel}, one can construct a field
subalgebra which commutes with the constraints (\ref{2.45}) and is
isomorphic to the field subalgebra of the vector Schwinger model
(QED$_2$), but
does not belong to the intrinsic field algebra of the anomalous
chiral model.

The physical Hilbert
space is constrained by the requirement, that the operators
(\ref{2.44}) annihilate the physical states. The physical fermion field
$\psi_2$ is expressed in terms of the free
field according to eqs. (2.6), (2.7),(2.26) and (2.27) as

$$ \psi_2 = U^{-1}\beta^{-1}{}\tilde{\Sigma}\psi^{(0)}_2 = : e^{2i\sqrt{\pi}
\left( \sigma-\eta+\varphi\right) }:\psi^{(0)}_2\,.\eqno (4.1) $$

\noindent We bosonize the free fermion in terms of $\phi$ and consider the
physical composite
operator $\psi_2\psi^{(0) \dagger}_1$,

$$ : \psi_2\psi^{(0) \dagger}_1 : =
:e^{2i\sqrt{\pi}\left(\sigma-\eta+\varphi+\phi\right)}:
\,.\eqno (4.2) $$

\noindent Defining the new fields $\varphi^\prime=\phi(x^+)+
\varphi(x^-)$ and $\phi^\prime=\varphi(x^+)+\phi(x^-)$ we see
that, in the case $a = 2$, the field  $\varphi^\prime$ decouples from
the longitudinal current (2.37) and there is a broader
class of operators belonging to $\Im^e$ that satisfy the constraint
conditions. The composite operator
(4.2) can be factorized as the product of two ``physical''
exponential operators that separately commute with the constraint

$$ : \psi_2 \psi^{(0) \ast}_1 : =
:e^{2i\sqrt{\pi}\left(\sigma-\eta+\varphi^\prime \right)}:
:e^{2i\sqrt{\pi} \phi^\prime}:\,.\eqno (4.3) $$

\noindent The first exponential factor appearing in (4.3) is the
Lowenstein-Swieca solution of the Schwinger model
leading for instance to $\theta$-vacua parametrization \cite{LS}. As stressed
in  \cite{Bel} the operator (4.3) cannot be reduced by extracting the
exponential of the field $\phi^\prime$, since the corresponding exponential
operator separatelly does
not belongs to the intrinsic field algebra of the
anomalous chiral model. Although the operator $ :\exp \{2 i \sqrt \pi
\phi^\prime \}: \in \Im^e $ commutes with the constaint (\ref{2.45}), it cannot
be defined on
the Hilbert space ${\cal H}$. This follows from the fact that some
charges carried by the field $\phi^\prime$ get trivialized in the
restriction from ${\cal H}^e$ to ${\cal H}$ \cite{MPS,Bel}. In this way, the
equivalence of vector Schwinger model and chiral Schwinger model
with $ a = 2 $ and the need for a $\theta$-vacuum parametrization in the
anomalous model, as suggested in ref.\cite{CW}, cannot be established in
terms of
the intrinsic field algebra and are consequence of an improper factorization
of the completion of states \cite{Bel}.

Returning to the nonabelian case the same analysis applies to the
$U(1)$ piece of the $U(N)$ model. To this end let us
separate the fields into $U(1) \times SU(N)$ \cite{FUR,GEP}. Within
our present decoupling method we consider the Wess-Zumino-Witten fields
now taking values on $U(N)$. Instead of
(2.30) the fields are expressed in terms of an $U(1)$ factor and
$SU(N)$-valued field operators as:

$$ \hat\beta=e^{-2i\sqrt{\pi\over N}\sigma}\beta,\quad
\hat\Sigma=e^{-2i\sqrt{\pi\over N}\eta} \tilde \Sigma,
\quad \hat U=e^{-2i\sqrt{\pi\over N}\varphi}U\,.\eqno (4.4) $$

All the decoupling and the analysis of the previous sections are the same
with the $C_V$ factors appearing in the actions multiplying
only  the $SU(N)$ WZW fields.  Factoring out the $U(1)$ dependence of
the
free fermion in terms of the scalar field $\phi$  \cite{WZW,Affleck},
instead of (4.3) we obtain

$$ \hat\psi_2 \hat\psi_1^{(0)\dagger} =
:e^{2i\sqrt{\frac{\pi}{N}}\left(\sigma-\eta+\varphi^\prime \right)}:
:e^{2i\sqrt{\frac{\pi}{N}}\,\phi^\prime }:\,\Biggl [\,
 U^{-1}\beta^{-1}\tilde \Sigma \Bigl (
\psi^{(0)}_2\psi^{(0)\dagger}_1 \Bigr )
\Biggr ]\,.\eqno (4.5) $$

\noindent Once again the first exponential factor, exluding the massive
field $\sigma$, is a spurious operator with zero scale dimension and
spin that generates constant Wightman functions \cite{BSRS}. It is
tempting to extract
from the operator (4.5) the dependence on
the field $\phi^\prime$, since for $a = 2$ it commutes with the BRST
constraints. With this procedure one would erroneously conclude
for the need of the $\theta$-vacua parametrization in the $U(N)$ model. Within
the formulation in terms of the intrinsic field
algebra generated from the irreducible set of field operators \cite{Bel}, the
spurious operators and the operator $:\exp \{ 2 i \sqrt \pi \phi^\prime \}:$ do
not exist separately and cannot be defined on ${\cal H}$.

It would be interesting to generalize the method here used to the case
$ a \neq 2 $, in order to seek a transformation decoupling the fields in
this case and, eventually obtain a complete decoupled formulation for
the chiral anomalous QCD$_2$.

\section*{Acknowledgement}
One of the authors (K.D.R) would like to thank the Physics Department
of the ``Universidade Federal Fluminense'' and of the University of
Stellenbosch for their kind hospitality and financial support,
 which made this collaboration possible.


\section*{Appendix}
\renewcommand{\theequation}{A.\arabic{equation}}
\setcounter{equation}0

In this appendix we show that the Lowenstein-Swieca type constraints are the
only ones to be imposed, and that the additional BRST symmetry relating the
conformal to the massive $\beta$ sector in the non-local formulation
implies no restriction on the physical states. We do this in the context of the
Schwinger model, where the results may be familiar to the reader.
Our conclusions will however equally apply to the chiral Schwinger model
and chiral $QCD_2$. For future reference we make the discussion
self-contained, at
the expense of some redundancy.

To begin with we recall the general procedure for identifying the
BRST symmetry and charges associated with a change of variables.
Consider the generic partition function

\begin{equation}
\label{partf}
Z = \int \, [d \phi] \, e^{i \, S \, [\phi]}\,,
\end{equation}
where $\phi$ stands for a generic set of fields.
We now consider a change of variables $\phi
\, \to \, \sigma$ given by $\phi = f \, (\sigma)\,$.
We implement the change of variable $\phi \, \to \, \sigma$ by
introducing in (\ref{partf}) the identity

\begin{equation}
\label{ident}
1 = \int \, [d \sigma] \, \det \, \left(\frac{\delta \, f}{\delta \,
\sigma}\right) \, \delta \, (\phi - f \, (\sigma)) \; .
\end{equation}
Using the Fourier representation of the delta--function, and representing
the functional determinant in
terms of ghosts, we arrive at the partition function

\begin{equation}
\label{partfex}
Z = \int \, [d \phi] \, [d \rho] \, \int \, [d b] \, [d c] \,
 e^{i \, S \, [\phi]} \cdot e^{i \, \int  \rho \,
(\phi - f \, (\sigma)) + i \, \int  b \,
 \frac{\delta \, f}{\delta \, \sigma} \, c}\,,
\end{equation}
where summation over the indices labelling the fields is understood.
As is well known \cite{Ba} there is a
BRST symmetry associated with the change of variable which is readily
read off from (\ref{partfex}):

\bear\label{brs1}
\delta \, \rho &=& \delta \, \phi = \delta \, c = 0\nonumber\\[3mm]
\delta \, \sigma &=& c ,\quad  \delta \, b = \rho
\ear
This symmetry is off--shell nilpotent.  In terms of the graded variation
$\delta$, the effective action
${\cal S}_1$ in (\ref{partfex}) can be written as $S \, [\phi]$ plus a
$\delta$ exact term: ${\cal S}_1 = S \, [\phi]
+ \delta \, (b \, (\phi - f \, (\sigma)))$.  Hence, in order to have
 equivalence of the extended description
(\ref{partfex}) with the original one as given by (\ref{partf}),
we must require that the transformation (\ref{brs1}) be a symmetry
of the physical states, and of any operator acting on them.

To implement the change of variables we integrate over $\rho$ and $\phi$
leaving us with the extended action

\begin{equation}
\label{exaction}
{\cal S}_1 = S \, [f \, (\sigma)] + \int \, b \, \frac{\delta \, f}{\delta \,
\sigma} \, c \; ,
\end{equation}
and the transformation (\ref{brs1}) now reads

\begin{equation}
\label{brst2}
\delta \, \sigma = c \, , \quad \delta \, c = 0 \, , \quad \delta \,
 b = - \left(\frac{\delta \, S \,
[\phi]}{\delta \, \phi}\right)_{\phi = f \, (\sigma)} \; .
\end{equation}
One readily checks that this transformation is a symmetry of (\ref{exaction}).
This symmetry is required to be a symmetry of the physical states and operators.

Now, the equation of motion for $\sigma$ reads
\be\label{eqmotion}
\left [\left (\frac{\delta S}{\delta \phi_{\alpha}}\right
)_{\phi=f(\sigma)}\right ]
 \left (\frac{\delta f_{\alpha}}{\delta \sigma_{\beta}}\right ) = 0.
\ee
Hence, as long as the transformation $\phi=f(\sigma)$
 is invertible (one-to-one mapping),
$\delta b = 0$ on shell, so that the BRST symmetry implies no constraint on the
states.
This shows, that non trivial constraints on the states are associated with
mappings which are not one-to one. In that case the BRST
symmetry of the states insures that the formal identity introduced in
order to realize the desired change of variables does indeed act as
the identity in the space of BRST invariant functionals, also in the
case where the mapping is not one-to-one.

In \cite{AA1,CRS} the transition from the fermionic formulation of $QCD_2$
to the so called "non-local" formulation involved two sequential
changes of variable. In reality, however,
only one set of transformations is involved. This becomes clear if we
 choose to write
from the start the Yang-Mills Lagrangian of $QCD_2$ in a
"first order" form by making use of the identity (compare with (\ref{2.21}))
\be\label{firstorder}
e^{i\int \frac{1}{2} tr{F^2}}=\int d[E] e^{i\int tr[\frac{1}{2} E^2 - EF]} \ee
where
\be\label{F}
F=\frac{1}{2} \epsilon^{\mu\nu}F_{\mu\nu}\ee
and then performing the change of variable (\ref{2.3}) and (\ref{2.4}), i.e
\bear\label{AUV}
eA_+ &=& U^{-1}i\partial_+ U, \nonumber \\
eA_- &=& Vi\partial_- V^{-1}, \nonumber \\
\partial_+ E &=& e\frac{(1+c_V)}{2\pi}\beta^{-1} i\partial_+ \beta \ear
The above mappings are evidently not one-to-one,
 so that we can expect non-trivial BRST constraints. As we wish to show now,
there actually exist
only two nontrivial such constraints associated with the first two
transformations involving the gauge field. Let us illustrate this
for the case of the Schwinger model.

With the parametrization
\bear\label{param}
U = e^{-ie\mu},\quad
V = e^{ie\nu}, \quad
\beta = e^{-i2\sqrt{\pi}\sigma}
\ear
the transformations (\ref{AUV}) read
\bear\label{abparam}
A_+ = \partial_+ \mu, \quad
A_- = \partial_- \nu, \quad
\partial_+E = m\partial_+ \sigma
\ear
where we have set $m = \frac{e}{\sqrt{\pi}}$. It is clear that the
above transformations do not represent a one-to-one mapping.

Corresponding to (\ref{ident}) we realize the change of variables
by introducing the identity
\bear\label{identity}
1= \int [d\rho][d\bar\rho][d\hat\rho] \int [d\mu][d\nu] \int [d(ghosts)]
e^{i\int \bar\rho (A_+ - \partial_+ \mu) + i\int b_- i\partial_+
c_-}\nonumber\\
\times e^{i\int \rho (A_- - \partial_- \nu) + i\int b_+ i\partial_- c_+}
e^{i\int \hat\rho(\partial_+ E - m\partial_+ \sigma) +
i\int \hat b_- i\partial_+ \hat c_-}
\ear
If zero modes of the light-cone derivatives are present, then (\ref{identity})
is not truly an identity. In this case BRST symmetries have to be
respected by the correlators, in order to turn (\ref{identity}) into an
identity on the space of BRST invariant functionals. The integral
in (\ref{identity}) exhibits the following BRST symmetries
\bear\label{BRST2}
(A) \quad \delta \bar\rho &=& \delta c_- = 0,\quad
\delta\mu = ic_- ,\quad
\delta b_- = \bar\rho  \cr
(B) \quad \delta\rho &=& \delta c_+ = 0,\quad
\delta\nu = ic_+,\quad
\delta b_+ = \rho \nonumber\\
(C)\quad \delta\hat\rho &=& \delta \hat c_- = 0,\quad
 \delta\sigma = i\hat c_-,\quad
\delta \hat b_- = m\hat\rho
\ear
All these transformations are off-shell nilpotent and commute with each
other. The
resulting effective action can be written as the sum
of the original fermionic action plus three BRST exact parts:
\be
S_{eff}[A,E,\psi^\dagger,\psi] + \Delta_A + \Delta_B + \Delta_C
\ee
where
\bear\label{Delta}
\Delta_A &=& \delta_A(b_-(A_+ - \partial_+\mu)), \cr
\Delta_B &=& \delta_B(b_+(A_- - \partial_-\nu)), \cr
\Delta_C &=& \delta_C(\hat b_-(\partial_+E - m\partial_+\sigma))
\ear
Hence the three BRST symmetries are to be imposed on the states in order to
insure that (\ref{identity}) really acts like the identity.

We now show that under
suitable asymptotic conditions the third of these BRST symmetries implies
 no restriction on shell.  To this end
we integrate over $A_{\pm}, \rho, \bar\rho$, as well as $\hat\rho$ ,$E$.
The integration over the first set of variables is unproblematic, and yields
for the corresponding $b$-ghost transformations,
\bear\label{BRST3}
\delta_A b_-&=& \frac{1}{2} \partial_- E - e\psi_1^{\dagger}\psi_1 \nonumber\\
\delta_B b_+ &=& -\frac{1}{2} \partial_+ E - e\psi_2^{\dagger}\psi_2.
\ear
The integration over $E$ requires a special consideration, since it
involves the light-cone derivative of $E$, implying an invariance of the
effective action under translations of $E$ by functions involving the
light-cone coordinate $x^+$ only (zero modes of $\partial_+$). 
From the equation of motion (see (\ref{firstorder})), $E=F$, and $F=\Box\chi$,
with $\chi$ the gauge invariant combination
$\chi = \frac{1}{2}(\mu-\nu)$, it follows that $E$ does not contain
zero modes. Hence $E$ is uniquely determined as a function of
$\sigma : E(x^+,x^-)=\sigma(x^+,x^-) - \sigma(-\infty,x^-)$, and corresponds
to the inverse of $\partial_+$ being defined by
$<x|{\partial_+}^{-1}|y> = \theta (x^+ - y^+)$.
Excluding zero modes in $\sigma$
then implies the (in the abelian case trivially) one-to-one mapping
$E = m\sigma$. Under these conditions
the transformation for the $b$-ghosts read:
\bear\label{BRST4}
\delta_A b_-&=& \frac{m}{2} \partial_-\sigma - e\psi_1^{\dagger}\psi_1 \cr
\delta_B b_+ &=& -\frac{m}{2} \partial_+\sigma - e\psi_2^{\dagger}\psi_2, \cr
\delta_C \hat b_-&=& m\partial_+^{-1} (m\sigma - \Box \chi)
\ear
We now perform the chiral transformation
\be\label{chiral}
\psi_1 = U^{-1}\psi^{(0)}_1, \quad \psi_2 = V\psi^{(0)}_2
\ee
Taking account of the usual chiral anomaly, we have under
this change of variable
\bear\label{fcurrents}
e\psi_1^{\dagger} \psi_1 &=&
e\psi^{\dagger (0)}_1 \psi^{(0)}_1 - \frac{1}{2}m\partial_-(m\chi) \nonumber\\
e\psi_2^{\dagger} \psi_2 &=&
 e\psi^{\dagger (0)}_2 \psi^{(0)}_2 + \frac{1}{2}m\partial_+(m\chi) . \ear
The free fermionic bilinears admit the familiar \cite{LS,AAR}
bosonic representation
\bear\label{fermions}
\psi^{\dagger (0)}_1 \partial_+\psi^{(0)}_1 +
\psi^{\dagger (0)}_2 \partial_-\psi^{(0)}_2 =
\frac{1}{2}\partial_{\mu}\phi \partial^{\mu}\phi \nonumber\\
\psi^{\dagger (0)}_1 \psi^{(0)}_1 = -\frac{1}{2\sqrt{\pi}}\partial_-\phi, \quad
\psi^{\dagger (0)}_1 \psi^{(0)}_1 =  \frac{1}{2\sqrt{\pi}}\partial_+\phi
\ear
In the partition function the chiral transformation (\ref{chiral})
contributes an
anomalous term $\frac{m^2}{2}\chi\Box\chi$ to the action.

The field $\chi$ contains zero mass modes, which are disentangled by
defining a new (zero mass) field $\eta = m\chi + \sigma$. Summarizing
our results for $S_{eff}$ and the BRST transformations we have
\be\label{Seff}
S_{eff} = \int
[-\frac{1}{2} [\phi\Box\phi -\eta\Box\eta +\sigma(\Box + m^2)\sigma ]
+\int [b_+i\partial_+ c_+ + b_+i\partial_- c_+ + \hat b_- i\partial_+ \hat
c_- ]\ee
as well as (we have rescaled the transformation laws by dividing through by $m$)
\bear\label{BRST1}
\delta_A b_- &=& \partial_-(\eta + \phi), \quad
\delta_A\eta = -\delta_A\phi = ic_- \nonumber\\
\delta_B b_+ &=& -\partial_+(\eta + \phi), \quad
\delta_B\eta = -\delta_B\phi = ic_+ \nonumber\\
\delta_C \hat b_- &=& \partial_+^{-1} ((\Box + m^2)\sigma - \Box\eta), \quad
\delta_C\eta = \delta_C\sigma = i\hat c_- .
\ear
with all the remaining variations vanishing.
In the usual case, BRST constraints correspond to the requirement that
$\delta( b-ghost) = 0$ on the states. We see that in the first two cases this
leads to the familiar Lowenstein-Swieca conditions defining the physical
Hilbert space. On the other hand, because of the existence of the inverse
of the operator $\partial_+$ as discussed before, $\delta_C \hat b_- = 0$
is identically satisfied on mass shell, and thus represents no additional
 constraint on the states. This is completely in line with what is known
about the Schwinger model, and is consistent with the observation that
in the case of the Schwinger model the change of variable $E \to \sigma$
was superfluous, since by the simple change of
variable $\eta = m\chi + \frac{1}{m} E$ one already succeeded in decoupling
 the massive ($E$) and massless ($\eta$) degrees of freedom.
 This is no longer so in the case of $QCD_2$, in which case
the change of variable (\ref{AUV}) required to achieve this decoupling
takes one from the Lie-algebra valued field $E$ to the group-valued
field $\beta$:
\be\label{Ebeta}
\beta^{-1}(x^+,x^-) = P e^{\frac{i}{\lambda}
\int_{-\infty}^{x^+} dy^+ \partial_+ E} \beta^{-1}(-\infty, x^-)
\ee
By requiring $\beta(-\infty,x^-) = 1$ we again exclude zero modes
and the mapping becomes one-to-one.

Our previous considerations indicate that only the conformal sector of $QCD_2$
is constrained by BRST conditions. This is in conflict with the claims made
in \cite{CR}.
To gain further insight into the problem let us review in the context of the
Schwinger model the steps followed in \cite{CR},
which led to a phase space representation
of $\delta_C\hat b_-$. They correspond to rewriting $\frac{1}{2}m^2\sigma$ in
(\ref{Seff}) as follows:
\be\label{massterm}
-\int \frac{m^2}{2}\sigma^2 \to
\int \left[ \frac{1}{2}(\partial_+ B - m\sigma)^2
-\frac{m^2}{2}\sigma^2\right]
=\int \left[ \frac{1}{2} (\partial_+ B)^2 - m\sigma\partial_+ B\right].
\ee
Assuming $\partial_+B$ to have no zero modes, we obtain upon integrating once
the equation of motion $\partial_+^2 B = m\partial_+ \sigma$
(since $\sigma$ is a massive field, we may set $\sigma(-\infty,x^+) = 0$)$^4$

\be\label{Bequation}
\partial_+ B = m\sigma.
\ee
Consider once more the operator
$\hat\Omega_- = \partial_+^{-1}\left[(\Box + m^2)\sigma - \Box\eta\right]$
appearing on the right hand side of (\ref{BRST1}). 
Noting that $\partial_+^{-1}$
stands for $\int_{-\infty}^{x^+}$ and making use of (\ref{Bequation}) we obtain
\be\label{Op}
\hat\Omega_- = [\partial_-(\sigma - \eta) + mB]
-[\partial_-(\sigma - \eta) + mB]_{x^+ = -\infty}
\ee
Hence $\hat\Omega_-=0$ implies
\be\label{localeq}
\partial_-(\sigma - \eta) + mB = 0.
\ee
Hence for this equation to be satisfied,
 we must allow for a chiral zero mode $\partial_-\eta$ in $B$. This zero mode
is right moving so that $\partial_+B$ has no zero mode in agreement with
the assumption made above.
In phase space, eq. (\ref{localeq}) reads (see (\ref{Seff}) and (\ref{massterm}))
\be\label{3dconstraint}
\pi_{\sigma} + \pi_{\eta} - \partial_1\sigma + \partial_1\eta + mB=0.
\ee
From (\ref{massterm}) one obtains for the momentum conjugate to $B$
\be\label{Bconjmom}
\pi_B = \partial_+ B - m\sigma.
\ee
Hence the equation of motion (\ref{Bequation}) actually
represents the constraint $\pi_B = 0$!
Since this constraint does not commute with $\hat\Omega_-= 0 $ we implement
these two
constraints strongly. Setting $\Phi_1 = \hat\Omega_-$ and $\Phi_2 = \pi_B$
we have
\be\label{Qmatrix}
\{\Phi_i(x) , \Phi_j(y)\} = m\epsilon_{ij} \delta (x^1 - y^1).
\ee
Because of the $\epsilon$-tensor one finds that the Poisson brackets of the
constraints
$\Omega_{\pm}$ remain unchanged with respect to the new Dirac-brackets.
 The whole procedure thus shows that the constraint $\hat\Omega_- = 0$, now
implemented
strongly, just serves to determine the field $B$ as a function of the
 remaining
fields.  This is in accordance with the point of view of ref. \cite{CRS} and
shows once
more that the third BRST symmetry in (\ref{BRST1}) does not imply a constraint
on the states.  The constraints thus only operate in the conformal (massless)
sector.

The non-abelian case actually corresponds to writing (\ref{Seff})
in the partially integrated form
\be\label{nabmassterm}
-\int \frac{m^2}{2}\sigma^2 \to
\int \left[ \frac{1}{2}(\partial_+ B - m\sigma)^2
-\frac{m^2}{2}\sigma^2\right]
=\int \left[ \frac{1}{2} (\partial_+ B)^2 - m\sigma\partial_+ B\right].
 \ee
Following the same argumentation as above, we are lead to the two constraints
\bear\label{constraints}
\Phi_1 &=& \pi_{\sigma} + \pi_{\eta} - \partial_1(\sigma - \eta)=0 \nonumber\\
\Phi_2 &=&\pi_B - m\sigma=0.
\ear
We again have the property (\ref{Qmatrix}), showing that these constraints
are to be implemented strongly. In terms of the corresponding
Dirac brackets, the constraints $\Omega_{\pm}\approx 0$ are first class, and
are thus to be implemented on the states.

In the non abelian case the constraints $\Phi_2\approx 0$ take a
non-local form in phase space, so that an implementation
a la Dirac is not possible. The general ideas however remain valid.
It is clear that the same type of argument also applies to
chiral $QCD_2$.

\newpage

{\bf FOOTNOTES}

\bigskip\noindent
1) Our conventions are: $A_\pm = A_0 \pm A_1\,, \partial_{\pm} =
\partial_0\pm\partial_1 \,, \epsilon^{01} = 1$. We follow
here the notation of refs. \cite{CRS}, \cite{CR}.\\
\noindent
2) See also the abelian case discussed in Ref.\cite{Bel}.\\
\noindent
3) Note the -- sign on the r.h.side. Indeed, our definitions (\ref{3.5})-(\ref{3.6}) differ from the conventional ones by a -- sign.\\
\noindent
4) Instead of (\ref{massterm}) we could have first considered the
replacement 
\be
-\int \frac{m^2}{2}\sigma^2 \to
\int \left[ \frac{1}{2}(G - m\sigma)^2 - \frac{1}{2}m^2\sigma^2\right]
\nonumber\\
\ee
and then have made the change of variable $G=\partial_+B$. Following the 
method of (\cite{Ba}), eq. (\ref{Bequation}) would then emerge as the BRST
constraint associated with this change of variable.

\end{document}